# Validation of automatic passenger counting: introducing the t-test-induced equivalence test

**Michael Siebert[1] · David Ellenberger[1,2]**

## Abstract

Automatic passenger counting (APC) in public transport has been introduced in the 1970s and has been rapidly emerging in recent years. Still, real-world applications continue to face events that are difficult to classify. The induced imprecision needs to be handled as statistical noise and thus methods have been defined to ensure that measurement errors do not exceed certain bounds. Various recommendations for such an APC validation have been made to establish criteria that limit the bias and the variability of the measurement errors. In those works, the misinterpretation of non-significance in statistical hypothesis tests for the detection of differences (e.g. Student's t-test) proves to be prevalent, although existing methods which were developed under the term *equivalence testing* in biostatistics (i.e. bioequivalence trials, Schuirmann in J Pharmacokinet Pharmacodyn 15(6):657–680, 1987) would be appropriate instead. This heavily affects the calibration and validation process of APC systems and has been the reason for unexpected results when the sample sizes were not suitably chosen: Large sample sizes were assumed to improve the assessment of systematic measurement errors of the devices from a user's perspective as well as from a manufacturers perspective, but the regular t-test fails to achieve that. We introduce a variant of the t-test, the revised t-test, which addresses both type I and type II errors appropriately and allows a comprehensible transition from the long-established t-test in a widely used industrial recommendation. This test is appealing, but still it is susceptible to numerical instability. Finally, we analytically reformulate it as a numerically stable equivalence test, which is thus easier to use. Our results therefore allow to *induce* an equivalence test from a t-test and increase the comparability of both tests, especially for decision makers.

✉ David Ellenberger
   david.ellenberger@interautomation.de

1   Interautomation Deutschland GmbH, Hauptstrasse 56-60, 13158 Berlin, Germany

2   Department of Medical Statistics, University Medical Center Göttingen, Humboldtallee 32, Göttingen 37073, Germany

Springer

## Introduction

Assessment of passenger counts is of paramount importance for public transport agencies in order to plan, manage and evaluate their transit service. Application covers many topics, for example short- and long-term forecasting, optimizing passenger behaviour and daily operations, or sharing of revenue among operators. Issues of passenger demand have a long-lasting history (see e.g. Kraft and Wohl 1968). In recent years, modelling of passenger counts has emerged rapidly due to the availability of large-scale automatic data collection. Data on (automatic) passenger counts has direct impact on both the revenue generated by ticket sales as well as state subsidies of public transport companies within one unified ticketing system. To illustrate, in Berlin, buses and underground trains are operated by the BVG, while a complementary rapid transit system is operated by S-Bahn Berlin GmbH. Public transport services in Berlin (and Brandenburg) are provided by around 40 companies organized in the Transport and Tariff Association of Berlin and Brandenburg (VBB), which provides operator-spanning tickets, e.g. single day or monthly passes. Excluding subsidies, the revenue from the ticket sales alone has been around 1.4 billion EUR in 2017[1]. Revenue magnitudes in the billions are common in public transport (Armstrong and Meissner 2010) and regularly need to be shared among different operators on the basis of (automatic) passenger data. APC systems have evolved considerably within the past 40 years. Passenger flow data can be acquired with high accuracy outperforming manual ride checkers (Hodges 1985; Hwang et al. 2006). Devices that operate on 3D image streams are the industrial state-of-the-art technology. Latest generation devices offer an accuracy of around 99% (iris 2018; Hella Aglaia 2018) and technical progress is ongoing. As all measurement devices, APC systems are susceptible to error. For the comparison of counting precision between different sensors, objective statistical criteria are therefore required. These criteria are not only needed for comparisons between APC systems but also decision-making processes rely on high accuracy APC data (Furth et al. 2005). Upraising usage of APC systems led to the formulation of some position criteria to ensure validity and reliability. The term *APC validation* for this type of quality control was used by Strathman (1989) and some wider usage of validation concepts awoke since the early 2000s (see e.g. Kimpel et al. 2003; Strathman et al. 2005; Boyle 2008; Chu 2010; Köhler et al. 2015). For real-world APC validations the most relevant criterion is to ensure unbiasedness, i.e. the need to rule out that the APC system makes a relevant a systematic error. Especially regarding revenue sharing, for which APC count data is commonly used (Detig et al. 2014; Hagemann 2017; Nahverkehrs-praxis 2014; Verkehr & Technik 2016; VVS 2016; VMT 2010), this is crucial, since small errors—of whatever origin—can already have a large impact: to illustrate, companies with a shared ticketing system like the above-mentioned BVG or the S-Bahn Berlin GmbH have revenues, consisting of ticket sales and subsidies, of roughly one billion EUR each[2]. If one of these companies somehow systematically counted 1% too few and the other one 1% too many passengers and passenger counts accounted to only 10%[3] of the shared revenue, yet two million EUR would be distributed inappropriately—every year, for these two companies operating in the Berlin area alone. In Germany, it is prevalent that the tickets are sold by the transport association and revenue, as well as subsidies, are distributed among the transport companies (Beck 2011), which accounted for 12.8 billion EUR in 2017 (VDV 2018). Furthermore, such a revenue sharing scheme itself is currently associated with high costs, which were roughly one million EUR for the VBB in the year 2014 (Baum and Gaebler 2015).

One industrial recommendation regarding APC systems is central to tendering procedures in Germany and also advertised by manufacturers worldwide (Hella Aglaia 2018; iris 2018): the VDV 457 (Köhler et al. 2015), which regularly and in an unmodified form becomes part of transportation contracts, sometimes even in the latest, yet unreleased version that has "this is a pre-release" watermarked on every page. Due to the huge impact of the document, all change requests to the VDV 457 must be approved by a committee within the Association of German Transport Companies (*Verband Deutscher Verkehrsunternehmen*, VDV). Results presented in this manuscript are given as follows: In the second section we summarize and discuss the development in (automatic) passenger counting. The complete statistical model of APC measurements we introduce in the third section. In the fourth section we define and examine the *revised t-test*, which is an attempt to modify resp. extend the t-test to account for the type II error accordingly. There were two reasons for this approach: Firstly, the admission process based on the t-test was already established in the VDV 457 v2.0 and its predecessors, so we wanted to change as little as possible to make the impact of the changes foreseeable and manageable by decision makers. Secondly, being unopinionated was more likely to succeed than simply insisting to use a statistical test because it was popular in other fields like biostatistics. Subsequently, we illustrate that this newly introduced test generally suffers from numerical instability making the approach unsuitable for wide practical use. In the fifth section we introduce the equivalence test and in the sixth section we normalize the test criteria of both the revised t-test and the equivalence test to analytically see, that, after transposition of parameters, the tests are identical. This so-obtained *t-test-induced equivalence test* however is, due to only elementary calculations being made, generally not susceptible to numerical instability. We close with some concluding remarks and future prospects in the last section.

## APC development and current practice

Traditionally, but also nowadays, passenger counts are collected manually via passenger surveys or human ride checkers, which are both expensive and produce only small samples. Former, the passenger surveys, are possibly biased and unreliable (Attanucci et al. 1981). For latter, the manual counts by ride checkers, the accuracy is doubtable, since already the first-generation automatic counting systems have been regarded to be more accurate (Hodges 1985). Ride checking is often done by less qualified personnel with high turn-over rates and Furth et al. (2005) instead suggest the use of video cameras to increase accuracy and reliability. Today, automatic data collection (ADC) systems in public transport are classified into automatic vehicle location (AVL), automatic passenger counting (APC), and automatic fare collection (AFC) systems (Zhao et al. 2007). The AVL system provides data on the position and timetable adherence of the bus, metro, or train which needs to be merged with APC data (Furth et al. 2004; Strathman et al. 2005; Saavedra et al. 2011). AFC data is based on ticket sales, magnetic strip cards, or smart cards and has become popular since it is often easily available (Zhao et al. 2007; Lee and Hickman 2014). Still, it often only provides information on boarding but not alighting, generally underestimates actual passenger counts and may therefore be less accurate than APC data (see e.g. Wilson and Nuzzolo 2008; Chu 2010; Xue and Sun 2015).

---

[3] Based on Beck (2011): 10% of the contracts were net-cost contracts. The actual shares of the individual transport companies in the VBB are considered trade secrets of the individual transport companies and are thus not disclosed, not even to a request of the House of Representatives of Berlin (Baum and Gaebler 2015).

The first generation of APC systems was deployed in the 1970s (Attanucci and Vozzolo 1983) and usage increased in the following decades. Casey et al. (1998) reported that many local metropolitan transit agencies use APC systems and Strathman et al. (2005) reported increased rates in APC usage of over 445% within seven years. Today APC systems are used worldwide and have found their way into official documents, as the above-mentioned tendering procedures in Germany. In the United States, transit agencies using APC data have to submit a benchmarking and a maintenance plan for reporting to the FTA's National Transit Database (NTD) to be eligible for related grant programs (see e.g. Chu 2010).

A wide range of competing APC technologies has been developed. Detection methods include infrared light beam cells, passive infrared detectors, infrared cameras, stereoscopic video cameras, laser scanners, ultrasonic detectors, microwave radars, piezoelectric mats, switching mats, and also electronic weighing equipment (EWE) (Casey et al. 1998; Kuutti 2012; Kotz et al. 2015). Transit agencies usually mount one or multiple sensors to collect APC data in each door area of public transport vehicles like buses, trams, and trains. The number of boarding and alighting passengers are counted separately by converting 3D video streams (infrared beam break) or light barrier methods, which are the most commonly used technologies (Kotz et al. 2015). In recent years also weight based EWE approaches utilizing pressure measurements in the vehicle braking/air bag suspension system have emerged to estimate passenger numbers (Nielsen et al. 2014; Kotz et al. 2015). These relatively new approaches have proven to provide easy-to-acquire additional information since modern buses and powertrains are equipped with (intelligent) pressure sensors by default.

First assessment of APC validity, i.e. accuracy, date back to the 1980s when large-scale usage started in the United States and Canada (Hodges 1985). To assess APC systems several researchers used confidence intervals and tests for paired data to investigate whether any found bias is statistically significant. The most commonly used statistical test is the t-test (Strathman 1989; Kimpel et al. 2003; Köhler et al. 2015), but the nonparametric Wilcoxon test for paired data has also been used in automated data (Kuutti 2012). Handbooks for reporting to FTA's National Transit Database have adapted the t-test as well as the industrial recommendations for APC-buying transit agencies like Köhler et al. (2015). To our knowledge no t-test related APC criterion formulated so far controls the type II error of the statistical test. Some authors report concepts that resemble equivalence testing. Furth et al. (2006) states *"A less stringent test would allow a small degree of bias, say, 2% (partly in recognition that the 'true' count may itself contain errors); [...]"* which acknowledges the fact that almost no measurement in the real world will have an expected value of exactly zero. In a survey among transit agencies by Boyle (2008) on how they ensure that APC systems meet a specified level of accuracy it is reported *"Some [agencies] were more specific, for example, with a confidence level of 90% that the observations were within 10% of actual boardings and alightings."*, which is an early occurrence of an equivalence test concept. Conversely, Chu (2010) introduced an *"equivalency test"* for APC benchmarking, which however is not to be confused with the equivalence test but rather is the application of the objected t-test to average passenger trip lengths. Additional *adjustment factors* on the raw APC counts are given without defining any equivalency criteria, an issue this paper shall address properly.

Various alternative criteria exist also to the t-test to assess accuracy resp. unbiasedness on the one hand and precision resp. reliability on the other hand. Nielsen et al. (2014) also investigate absolute differences in addition to evaluate the bias when analysing a weight-based APC approach. Restrictions on the absolute deviation from zero also limit the variability of the APC system. Criteria specifically on the variance of the APC have been made indirectly through the error rate or more specifically through specifying the allowed distribution of errors, see e.g. criteria *b* and *c* in Köhler et al. (2015), appendix E in Furth et al. (2003), or Boyle (2008). To the best of our knowledge, the most comprehensive and maintained industrial recommendation on APC validation and usage is the above-mentioned *VDV Schrift 457* (Köhler et al. 2015). The document gives guidance on most relevant APC topics, including sampling and standardization of APC validation. One major aspect of APC validation is the demonstration of adequate APC accuracy regarding which Köhler et al. (2015) state for the approval process: for an APC system to pass the admission process, its *systematic error* has to be at most 1%, which is verified by (a variant of) the t-test. Worldwide, there are similar formulations for the validation and thus admission of APC systems, like Furth et al. (2006), Boyle (2008), or Chu (2010).

However, scepticism arose within the industry when seemingly good performing APC systems started to fail the test. In February 2015, with the help of a brute force algorithm, we constructed a *proof of concept* for a failed (APC) t-test: the error is almost zero, i.e. with 1000 (or arbitrary many more) boarding passengers the sample has three measurements, one with an error of one, the other two with an error of two passengers. In that case, the APC system fails the t-test. Such a proof of concept led the count precision workgroup (*Arbeitsgemeinschaft Zählgenauigkeit*) of the VDV to add the equivalence test with additional restrictions, as an *exceptional alternative test* alongside the t-test in the VDV 457 v2.0 release in June 2015 (Köhler et al. 2015) to account for APC systems with a low error standard deviation. Indeed, the above-mentioned proof of concept would now be accepted by the new, hybrid test, but as it turned out later, current or near future APC systems would not profit, since the parameter choice was too hard to pass. Further, a remark was added to the VDV 457 v2.0 that *"in the advent of technological advance and increased counting precision, the admission process is still subject to change"*: at that time, there was still little insight into why a seemingly suitable and popular statistical test exposed such a seemingly arbitrary behaviour and it was not entirely clear how the equivalence test compared to the long-established t-test.

Detailed investigations showed that the VDV 457 v2.0 t-test variant only accounted for the type I error, defined to be 5% to 10%, which is the risk for an APC systems manufacturer to fail the test with a system with having a systematic error of zero. In the t-test terminology, this parameter is known as statistical significance $\alpha$. Conversely, the type II error $\beta$ is the risk of an APC system with a systematic error greater than 1% to obtain admission, which is the complement to the *statistical power* $1 - \beta$. The type II error and thus the statistical power was neither accounted for in the sample size planning nor in the testing procedure. Through the sample size formula it was implicitly 50%, assuming the a priori estimated standard deviation was correct. Otherwise, the higher the empirical standard deviation, the greater the type II error and vice versa. The statistical framework for APC validation and methods to address the current shortcomings are given in the following sections.

## Statistical model

Let $\Omega_0 = \{\omega_i\}, i = 1, \ldots, N$ be the statistical population of *stop door events* (SDE), which are used to summarize all boarding and alighting passengers at a single door during a vehicle (bus, tram, train) stop. Further, let $\Omega = \{\omega_{i_j}\}, i_j \in \{1, \ldots, N\}, j \in \{1, \ldots, n\}$ be a sample, which consists of either randomly or structurally selected SDE (e.g. by a given sampling plan). The size of the statistical population $N$ may be considered as the number of all SDE over the relevant time period, which is typically one or more years, so $N$ can be assumed to be unbounded and thus $N = \infty$. Let $n$ be the sample size and $M_i, i \in \{i, \ldots, n\}$ be the *manual count* and $K_i, i \in \{i, \ldots, n\}$ be the *automatic count* of boarding passengers made by the APC system. The manual count obtained by multiple ride checkers or favourably video camera information (Kimpel et al. 2003) is assumed to be a ground truth to compare against. Alighting passengers are counted as well and results apply analogously, but w.l.o.g. we only consider the boarding passengers. Let $\overline{M} = \frac{1}{n} \sum_{i=1}^{n} M_i$ be the average manual boarding passenger count. Similar to other authors [see e.g. appendix E in Furth et al. (2003), Furth et al. (2006), Nielsen et al. (2014), Köhler et al. (2015)] we consider the random variables

$$D_i := \frac{K_i - M_i}{\overline{M}}, \tag{1}$$

which we call *relative differences* being the difference of the automatically and manually counted boarding passengers relative to the average of the manually counted boarding passengers. The average $\overline{D} := \frac{1}{n} \sum_{i=1}^{n} D_i$ is the statistic of interest that is used in both the t-tests as well as the equivalence test. The expected value $\mu := E(\overline{D})$ is the *actual systematic error*[4] of an APC system (Furth et al. 2005), since it can systematically discriminate participants of the revenue sharing system or could also be referred to as *bias* of the measurement device, a term frequently used in APC accuracy evaluations (Strathman 1989; Kimpel et al. 2003; Furth et al. 2005; Chu 2010; Nielsen et al. 2014).

The criteria in each APC approval procedure are often reviewed by specially trusted authorities who are entitled to grant admission. They perform their own manual ride check, evaluate the criteria i.e. the statistical test, and either approve or reject the APC system. There are two conflicting interests that need to be dealt with: acquiring maximally accurate and reliable data on the one hand and approve a high number of APC in a fast and cost-efficient process on the other hand. Shortcomings of the first we will call *calibration* resp. *user risk* and shortcomings of the latter *manufacturer risk*. We attribute the user risk to public transportation companies and network authorities, who rely on accurate data, despite that the motivations for APC data collection might be more complex in the real world. The manufacturer risk relates directly to possible recourse claims and negative market reputation if the resp. APC system fails the admission. These two risks relate to the type I error and type II error of statistical tests. For the t-test, the hypotheses are

$$H_0: \quad \text{There is no systematic APC measurement error } (\mu = 0) \tag{2}$$

$$H_1: \quad \text{There is a systematic APC measurement error } (\mu \neq 0). \tag{3}$$

---

[4] Referred to as *distortion* in VDV 457 v2.1 (Köhler et al. 2018).

Let $v$ be the a priori estimated standard deviation, $\hat{v}$ the empirical standard deviation of the sample, $d_r$ the maximal allowed error (e.g. 1%), $\alpha_t$ the risk of falsely rejecting the null hypothesis $H_0$ (type I error, i.e. rejecting an APC system with an actual systematic error of zero) and $\beta_t$ the risk of falsely accepting the null hypothesis $H_0$ when a particular value of the alternative hypothesis $H_1$ is true (type II error, e.g. accepting an APC system with a systematic error of 1%) (see e.g. Guthrie 2010).

The sample size estimation for the t-test is given by

$$n_t = \left( z_{1-\beta_t} + z_{1-\alpha_t/2} \right)^2 \frac{v^2}{d_r^2}, \tag{4}$$

with $z_{(\cdot)}$ being the quantile function of the normal distribution and the test criterion as

$$|\overline{D}| \leq z_{1-\alpha_t/2} \frac{\hat{v}}{\sqrt{n_t}}. \tag{5}$$

## Revised t-test

Several discussions about post-hoc power adaptions for the t-test exist. A thorough discussion about those can be found in Hoenig and Heisey (2001). They argue that approaches referred to as *Observed Power*, *Detectable Effect Size*, or *Biologically Significant Effect Size* are *"flawed"*. For the latter approach, which is described in Cohen (1988), Hoenig and Heisey criticize the assumption that actual power is equal to the intended power and not updated according to experimental results (e.g. sampling variability). Addressing this, we investigate on procedures to control the (actual) type II error to assess non-presence of a crucial difference. Schuirmann (1987) initially referred to approaches of using a negative hypothesis test to make inference that *no inequivalence* was present as the *Power Approach*. Analogously to these thoughts, we will consider variations of the type I error $\alpha$ to make adaptations to the testing procedure and call this approach *post-hoc power calculation*. This was explicitly mentioned by Schuirmann but was not derived further by stating a lack of practical interest: *"In the case of the power approach, it is of course possible to carry out the test of the hypothesis of no difference at a level other than 0.05 and / or to require an estimated power other than 0.80, but this is virtually never done."* While this approach may not have been used in the world of pharmaceutics, it is of relevance for the validation of devices for automatic passenger count and likely other applications in industrial statistics. In general, as well in practice, after the data collection, the a priori estimated standard deviation and the empirical standard deviation differ to some extent and we strongly believe that it cannot be relied on that difference being negligible. Therefore, we want to ensure that the risk of the user (the type II error) does not exceed a prespecified level, which is usually 5%. So the only parameter to be adapted is the type I error $\alpha$, which is the risk of the device manufacturer. The appropriate $\hat{\alpha}_t$, the *revised significance*, can thus be determined by solving the equation

$$n_t \overset{!}{=} \left( z_{1-\beta_t} + z_{1-\hat{\alpha}_t/2} \right)^2 \frac{\hat{v}^2}{d_r^2}, \tag{6}$$

and thus

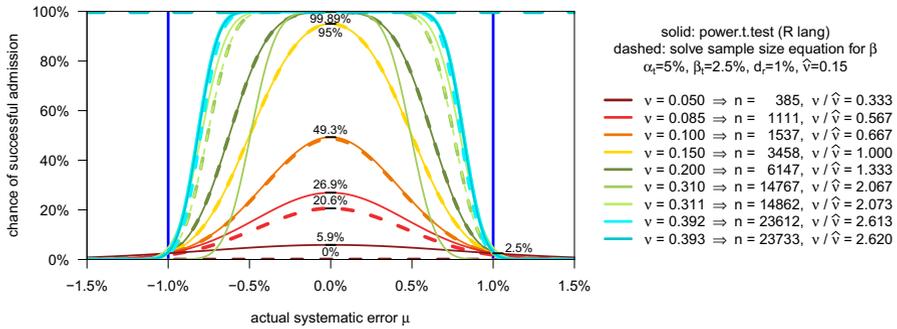

**Fig. 1** Chances (of an APC system) to pass dependent on the method to determine the revised significance $\hat{\alpha}_t$, which we have obtained from Eq. (7), denoted by dashed lines, and by using the function `power.t.test` from R Core Team (2018) denoted by the solid lines. Latter was our initial approach, which we illustrate here for the purpose of completeness. We notice that for far too low sample sizes (red and dark red lines) the former method is stricter. For the dark red dashed line, $n$ is below the lower bound from Eq. (9) and we thus assume the test to always fail, compare "t-test-induced equivalence test" section. For (too) large sample sizes numerical instabilities, which cannot be detected by the user through e.g. error messages, lead to sudden gaps in the function values (green lines for $\nu = 0.310$ and $\nu = 0.311$) for the `power.t.test`-variant, which relies on fixed-point iteration and has a history of unexpected convergence behaviour. The blue lines denote a practically reachable numeric limit when using Eq. (7): starting with $\nu \geq 0.393$ resp. $\nu/\hat{\nu} \geq 2.62$ all systems are always accepted. For the `power.t.test` approach, the light blue line is slightly above the dark blue one, so it has numerical problems for these values, too. Generally, all values of $\nu$ have to be seen w.r.t. the ratio $\nu/\hat{\nu}$, since these numerical effects can already occur with much smaller sample sizes. (Colour figure online)

$$\hat{\alpha}_t = 2\left[1 - z^{-1}\left(\left(z_{1-\beta_t} + z_{1-\alpha_t/2}\right)\frac{\nu}{\hat{\nu}} - z_{1-\beta_t}\right)\right]. \tag{7}$$

Note that $n_t$, by choice of the initial $\alpha_t$ and $\nu$, is fixed. If the actual sample size does not match the planned sample size, $n_t$ resp. $\nu$ needs to be adapted. Analogously to Eq. (5), we define the test criterion for the revised t-test:

$$|\overline{D}| \leq z_{1-\hat{\alpha}_t/2}\frac{\hat{\nu}}{\sqrt{n_t}}, \tag{8}$$

which, however, can yield a problematic behaviour in practice: First, the term $z_{1-\hat{\alpha}_t/2}$ is undefined for $\hat{\alpha}_t > 2$. Combined with Eq. (6), this induces a lower bound on $n_t$:

$$n_t \geq z_{1-\beta_t}^2 \frac{\hat{\nu}^2}{d_r^2}. \tag{9}$$

Second, $z(1) = \infty$ is the source of a different problem whenever $\nu/\hat{\nu}$ exceeds a certain critical value, which is 2.62 for $\alpha_t = 5\%$ and $\beta_t = 2.5\%$[5]. Therefore, this could be relevant in practice, since it yields $z_{1-\hat{\alpha}_t/2} = \infty$ due to rounding errors. In that case, the test criterion from Eq. (8) is always true and the test is thus always passed as illustrated in Fig. 1.

---

[5] Calculations were done on a 64 bit machine using e.g. R Core Team (2018) or Microsoft Excel.

## Equivalence testing

The equivalence test has its origin in the field of biostatistics. Often the term bioequivalence testing (e.g. Schuirmann 1987; Berger and Hsu 1996; Wellek 2010) is used. Bioequivalence tests are statistical tools that are commonly used to compare the performance of generic drugs with established drugs using several commonly accepted metrics of drug efficacy. The term *equivalence* between groups means that differences are within certain bounds, as opposed to complete equality. These bounds are application-specific and are usually to be chosen such that they are below any potential (clinically) relevant effect (Ennis and Ennis 2010). In many publications the problem was referred to as the *two one-sided tests* (TOSTs) procedure (Schuirmann 1987; Westlake 1981). TOST procedures were developed under various parametric assumptions and additionally distribution-free approaches exist (Wellek and Hampel 1999; Zhou et al. 2004).

Equivalence tests have begun making their way into psychological research (see e.g. Rogers et al. 1993) and natural sciences: Hatch (1996) applied it for testing in clinical biofeedback research. Parkhurst (2001) discussed the lack of usage of equivalence testing in biology studies and stated that equivalence tests improve the logic of significance testing when demonstrating similarity is important. Richter and Richter (2002) used equivalence testing in industrial applications and gave instructions on how to easily calculate it with basic spreadsheet computer programs. Applications also involved risk assessment (Newman and Strojan 1998), plant pathology (Garrett 1997; Robinson et al. 2005) ecological modelling (Robinson and Froese 2004) analytical chemistry (Limentani et al. 2005), pharmaceutical engineering (Schepers and Wätzig 2005), sensory and consumer research (Bi 2005), assessment of measurement agreement (Yi et al. 2008), sports sciences (Vanhelst et al. 2009), applications to microarray data (Qiu and Cui 2010), genetic epidemiology in the assessment of allelic associations (Gourraud 2011), and geography (Waldhoer and Heinzl 2011).

For the equivalence test, the hypotheses are (Schuirmann 1987; Julious 2004)

$$H_0 : \quad \text{There is a (relevant) systematic APC measurement error } (|\mu| \geq \Delta) \quad (10)$$

$$H_1 : \quad \text{There is no (relevant) systematic APC measurement error } (|\mu| < \Delta). \quad (11)$$

We define $\Delta$ to be the equivalence margin and the relevant errors for the equivalence test with $\alpha_e$ referring to (half) the risk of the user and $\beta_e$ to the risk of the device manufacturer. We will consider two-sided $1 - 2\alpha_e$ confidence intervals (symmetric around the mean) where $\alpha_e$ is often to be chosen 2.5%. The usage of $1 - \alpha_e$ confidence intervals is also possible (see e.g. Westlake 1981) but is used less frequently in the recent literature in this topic. Note that, by definition, the meanings of the $\alpha$ and $\beta$ are interchanged between the t-test and the equivalence criterion in referring to the risk of the user and to the risk of the manufacturer.

For the equivalence test sample size estimation exists (see e.g. Liu and Chow 1992; Julious 2004) similar to Eq. (4) of the t-test:

$$n_e = \left( z_{1-\beta_e/2} + z_{1-\alpha_e} \right)^2 \frac{\nu^2}{\Delta^2}. \quad (12)$$

We define the test criterion for the equivalence test

$$|\overline{D}| \leq \Delta - z_{1-\alpha_e} \frac{\widehat{v}}{\sqrt{n_e}}, \tag{13}$$

which is the formulation of the Two One-Sided Test Procedure for the crossover design in the case of limits that are symmetrical around zero (see Schuirmann 1987).

## t-test-induced equivalence test

An approach to compare the revised t-test to the equivalence test is to normalize and compare their test criteria from Eqs. (8) resp. (13) as well as their sample sizes formulas (4) resp. (12). By combining Eqs. (4), (6) and (8), we obtain a normalized test condition for the revised t-test:

$$
\begin{aligned}
|\overline{D}| \leq z_{1-\widehat{\alpha}_t/2} \frac{\widehat{v}}{\sqrt{n_t}} &= \left( \frac{v}{\widehat{v}} (z_{1-\beta_t} + z_{1-\alpha_t/2}) - z_{1-\beta_t} \right) \frac{\widehat{v}}{\sqrt{n_t}} \\
&= \left( \frac{v}{\widehat{v}} (z_{1-\beta_t} + z_{1-\alpha_t/2}) - z_{1-\beta_t} \right) \frac{\widehat{v}}{v} \frac{1}{z_{1-\beta_t} + z_{1-\alpha_t/2}} d_r \\
&= \left( 1 - \frac{\widehat{v}}{v} \frac{1}{1 + \frac{z_{1-\alpha_t/2}}{z_{1-\beta_t}}} \right) d_r.
\end{aligned}
\tag{14}
$$

Using Eqs. (12) and (13) we obtain

$$
\begin{aligned}
|\overline{D}| \leq \Delta - z_{1-\alpha_e} \frac{\widehat{v}}{\sqrt{n_e}} &= \Delta - z_{1-\alpha_e} \frac{\widehat{v}}{v} \left( \frac{1}{z_{1-\beta_e/2} + z_{1-\alpha_e}} \right)^2 \Delta \\
&= \left( 1 - \frac{\widehat{v}}{v} \frac{1}{1 + \frac{z_{1-\beta_e/2}}{z_{1-\alpha_e}}} \right) \Delta,
\end{aligned}
\tag{15}
$$

which resembles Eq. (14). If we now choose

$$\beta_e := \alpha_t, \ \ \alpha_e := \beta_t \ \ \text{and} \ \ \Delta := d_r \tag{16}$$

for Eqs. (12) and (15), they are identical to Eqs. (4) and (14). Therefore, the revised t-test *analytically is* an equivalence test, with error types swapped and an extended domain: Since only elementary calculations are made and there is no need to handle a varying quantile function $z_{(\cdot)}$, there is neither a lower bound as in Eq. (9), nor an upper bound due to numeric instability as illustrated in Fig. 1. We call an equivalence test with parameters chosen as in Eq. (16) a *t-test induced equivalence test*. For a visual comparison of the t-test and the equivalence test, see Fig. 2.

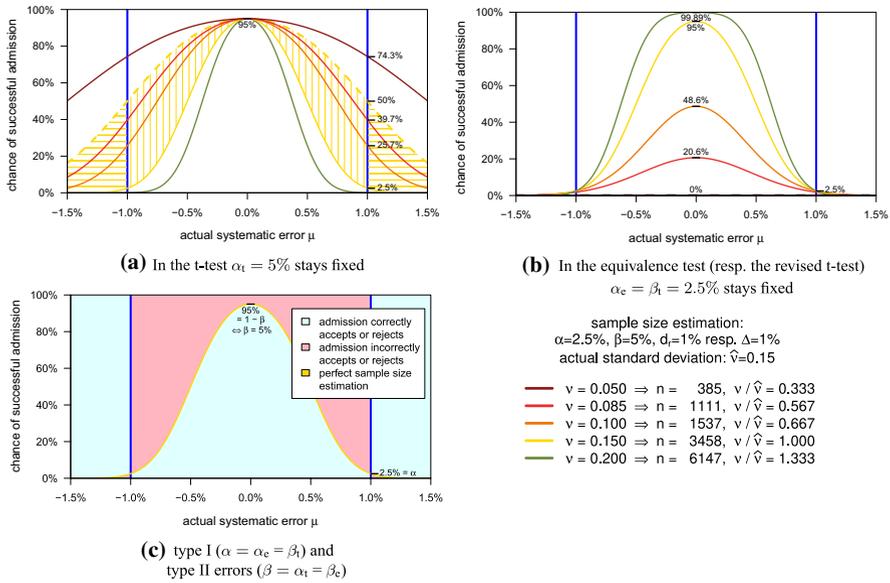

**Fig. 2** Chances of an APC system with an actual standard deviation of the (relative) errors equal to $\hat{v} = 0.15$ to pass **a** the t-test or **b** the revised t-test resp. equivalence test over the actual systematic error. Different lines denote different sample sizes obtained from different a priori choices of the standard deviation $v$. The golden, solid curve always represents a correctly estimated sample size, the green curve a sample which is too large and the reddish curves samples which are too small. The dashes in the dark red line in (**b**) denote the consequences of equation (9): only for the equivalence test, the outcome is defined for Eq. (13) and the test may be considered as always failed if $n = 385 < z_{1-\beta_t}^2 \cdot \hat{v}^2/d_r^2 = 864.33$. Using the original VDV 457 v2.0 with an (implicit) power of 50% yields the dashed golden curve in (**a**). The vertically striped areas are additionally *correctly* accepted, the horizontally striped areas are additionally *incorrectly* accepted. The thick blue lines denote the relative error of $d_r = 1\%$. For comparison: in (**c**) the incorrect decisions of a reference test are red, the correct decisions are coloured cyan. The reason for red areas to exist are economic considerations to limit the test costs: further increasing the sample size towards infinity would make the red areas disappear, at least for the revised t-test resp. the equivalence test. For the t-test, the areas with systematic error $\mu > 1\%$ and $\mu < -1\%$ remain blue, but the inner turns red. This behaviour is counterintuitive to the idea that the error of a statistic test goes to zero as the sample size goes to infinity. Note that, in the newly released VDV 457 v2.1, $\alpha/2$ is used at the place where we use $\alpha$ and therefore, our $\alpha = 2.5\%$ matches $\alpha = 5\%$ in Köhler et al. (2018). (Colour figure online)

## Conclusion

We illustrated that the t-test as a criterion for APC approval may exhibit undesirable properties, even as the sample size grows beyond a certain level. Further, we have shown that when trying to compensate this behaviour by using post-hoc power calculations with a revised t-test, issues of numeric stability and domain limitations arise. Finally, we have proven analytically that the t-test-induced equivalence test, being numerically stable with a practically unlimited domain, can supersede the revised t-test. The equivalence test is popular in various fields and, from a user's perspective, easier to apply than post-hoc power calculations. Our results thus not only apply to APC systems: *every* use of the t-test can now comfortably be reconsidered and—on demand—be replaced by a (t-test-induced) equivalence test.

Our work simplifies the decision-making process considerably, especially when it affects the worldwide revenue sharing in public transport, where there have been made 243 billion public transport journeys in the year of 2015 alone (UITP 2017). For this reason, a large German public transportation company, which was significantly involved in the creation of the original, t-test based recommendation, commissioned an additional, complementary experts report, which eventually confirmed our findings. With the release of the VDV 457 v2.1 in July 2018, our proposals have been accepted and the use of the equivalence test thus became the new recommendation for the validation of automatic passenger counting systems. Finally, we hope that long-lasting arguments within the industry about seemingly arbitrary admission results now end and also that our work will enable a broader audience to understand and profit from equivalence testing.

**Acknowledgements** This research is financially supported by the European Regional Development Fund.

**Authors' contributions** MS: Conceptual Guidance, Illustrations, Manuscript Writing and Editing. DE: Literature Search and Review, Manuscript Writing and Editing, Preparation of Supplementary Materials.

## Compliance with ethical standards

**Conflict of interest** Michael Siebert is an employee of Interautomation Deutschland GmbH. The submitted work does not pose a conflict of interest. David Ellenberger has been employed by University Medical Center Göttingen and Interautomation Deutschland GmbH during the time of research and manuscript preparation. None resulted in a conflict of interest.

## References

Armstrong, A., Meissner, J.: Railway Revenue Management: Overview and Models. Lancaster University Management School, Lancaster (2010)

Attanucci, J., Vozzolo, D.: Assessment of operational effectiveness, accuracy, and costs of automatic passenger counters. HS-037 821 (1983)

Attanucci, J., Burns, I., Wilson, N.: Bus Transit Monitoring Manual. Technology Sharing Program, Office of the Secretary of Transportation, Washington (1981)

Baum, A., Gaebler, C.: Ausgleichszahlungen im Rahmen der VBB-Einnahmeaufteilung (2015). http://www.stiftung-naturschutz.de/fileadmin/img/pdf/Kleine_Anfragen/S17-16190.pdf. Accessed 28 Mar 2019

Beck, A.: Barriers to entry in rail passenger services: empirical evidence for tendering procedures in Germany. Eur. J. Transp. Infrastruct. Res. **11**(1), 20–41 (2011)

Berger, R.L., Hsu, J.C.: Bioequivalence trials, intersection–union tests and equivalence confidence sets. Stat. Sci. **11**(4), 283–319 (1996)

Bi, J.: Similarity testing in sensory and consumer research. Food Qual. Preference **16**(2), 139–149 (2005)

Boyle, D.K.: Passenger Counting Systems (No. 77). Transportation Research Board, Washington (2008)

Casey, R.F., Labell, L.N., Carpenter, E.J., LoVecchio, J.A., Moniz, L., Ow, R.S., Royal, J.W., Schwenk, J.C., Schweiger, C.L., Marks, B.: Advanced public transportation systems: the state of the art. update '98. Technical report FTA-MA-26-7007-98-1, United States. Federal Transit Administration (1998)

Chu, X.: A guidebook for using automatic passenger counter data for National Transit Database (NTD) reporting (2010)

Cohen, J.: Statistical Power Analysis for the Behavioral Sciences. Lawrence Erlbaum Associates, Mahwah (1988)

Detig, R., Gilli, S., Axhausen, KW., Vitins, B.: Einnahmenaufteilung in Verkehrsverbunden, Bachelorarbeit Studiengang Geomatik und Planung (2014)

Ennis, D.M., Ennis, J.M.: Equivalence hypothesis testing. Food Qual. Preference **21**(3), 253–256 (2010)

Furth, P.G., Hemily, B., Muller, T., Strathman, J.G.: Uses of archived AVL-APC data to improve transit performance and management: review and potential. TCRP Web Document 23 (2003)

Furth, P., Muller, T., Strathman, J., Hemily, B.: Designing automated vehicle location systems for archived data analysis. Transp. Res. Record **1887**, 62–70 (2004)

Furth, P., Strathman, J., Hemily, B.: Part 4: marketing and fare policy: making automatic passenger counts mainstream: accuracy, balancing algorithms, and data structures. Transp. Res. Record **1927**, 205–216 (2005)

Furth, P.G., Hemily, B., Muller, T.H., Strathman, J.G.: Using archived AVL-APC data to improve transit performance and management. TCRP report (2006)

Garrett, K.: Use of statistical tests of equivalence (bioequivalence tests) in plant pathology. Phytopathology **87**(4), 372–374 (1997)

Gourraud, P.A.: When is the absence of evidence, evidence of absence? Genet. Epidemiol. **35**(6), 568–571 (2011)

Guthrie, W.F.: Sample sizes required. NIST/SEMATECH Engineering Statistics Handbook, NIST/SEMATECH (2010). http://www.itl.nist.gov/div898/handbook/prc/section2/prc222.htm. Accessed 28 Mar 2019

Hagemann, W.: Wo steht die automatische Fahrgastzählung? Der Nahverkehr 3 (2017)

Hatch, J.P.: Using statistical equivalence testing in clinical biofeedback research. Appl. Psychophysiol. Biofeedback **21**(2), 105–119 (1996)

Hella Aglaia: Public Transport: HELLA Aglaia People Sensing (2018). http://people-sensing.com/public-transport. Acessed 28 Mar 2019

Hodges, C.C.: Automatic passenger counter systems: the state of the practice. Technical report DOT-I-87-36 (398) (1985). https://rosap.ntl.bts.gov/view/dot/398

Hoenig, J.M., Heisey, D.M.: The abuse of power: the pervasive fallacy of power calculations for data analysis. Am. Stat. **55**(1), 19–24 (2001)

Hwang, M., Kemp, J., Lerner-Lam, E., Neuerburg, N., Okunieff, P.E.: Advanced public transportation systems: the state of the art update 2006. FTA report (2006) iris: Iris in Mannheim (2018). https://www.iris-sensing.com/de/referenzen/details/rhein-neckar-verkehr-gmbh-rnv/. Accessed 28 Mar 2019

Julious, S.A.: Sample sizes for clinical trials with normal data. Stat. Med. **23**(12), 1921–1986 (2004)

Kimpel, T., Strathman, J., Griffin, D., Callas, S., Gerhart, R.: Automatic passenger counter evaluation: implications for national transit database reporting. Transp. Res. Record **1835**, 93–100 (2003)

Köhler, S., Bobinger, S., Branick, R., Cerfontaine, B., Krogull, S., Luther, A., Ritschel, M., Schulze, M., Starck, M., Bruns, W.: Automatische Fahrgastzählsysteme: Handlungsempfehlungen zur Anwendung von AFZS im öffentlichen Personenverkehr, Version 2.0. VDV-Schrift 457, Verband Deutscher Verkehrsunternehmen (VDV), Köln (2015). https://gso.gbv.de/DB=2.1/PPNSET?PPN=832975419

Köhler, S., Bobinger, S., Branick, R., Cerfontaine, B., Krogull, S., Luther, A., Ritschel, M., Brunner, E., Ellenberger, D., Siebert, M., Schulze, M., Starck, M., Bruns, W.: Automatic Passenger Counting: Recommendations for the Application of APCSs within Public Transport and Regional Rail Transport, Version 2.1. VDV Recommendation 457, Verband Deutscher Verkehrsunternehmen (VDV), Köln (2018). https://www.beka-verlag.de/ebook-vdv-schrift-457-automatic-passenger-counting-systems-v2-1.html. Accessed 28 Mar 2019

Kotz, A.J., Kittelson, D.B., Northrop, W.F.: Novel vehicle mass-based automated passenger counter for transit applications. Transp. Res. Record **2536**, 37–43 (2015)

Kraft, G., Wohl, M.: New directions for passenger demand analysis and forecasting. RAND Corporation paper series P-3877 (1968)

Kuutti, J.: A test setup for comparison of people flow sensors (2012). http://urn.fi/URN:NBN:fi:aalto-201209203108

Lee, S.G., Hickman, M.: Trip purpose inference using automated fare collection data. Public Transp. **6**(1), 1–20 (2014)

Limentani, G.B., Ringo, M.C., Ye, F., Bergquist, M.L., McSorley, E.O.: Beyond the t-test: statistical equivalence testing. Anal. Chem. **77**(11), 221A–226A (2005)

Liu, J.P., Chow, S.C.: Sample size determination for the two one-sided tests procedure in bioequivalence. J. Pharmacokinet. Pharmacodyn. **20**(1), 101–104 (1992)

Nahverkehrs-praxis: Zehn Jahre Automatische Fahrgastzählsysteme (AFZS) im HVV (2014). https://www.nahverkehrspraxis.de/zeitschrift/heftarchiv/ausgaben/2014/112014. Accessed 28 Mar 2019

Newman, M.C., Strojan, C.: Risk Assessment: Logic and Measurement. CRC Press, Boca Raton (1998)

Nielsen, B.F., Frølich, L., Nielsen, O.A., Filges, D.: Estimating passenger numbers in trains using existing weighing capabilities. Transportmetrica A Transp. Sci. **10**(6), 502–517 (2014)

Parkhurst, D.F.: Statistical significance tests: equivalence and reverse tests should reduce misinterpretation. AIBS Bull. **51**(12), 1051–1057 (2001)

Qiu, J., Cui, X.: Evaluation of a statistical equivalence test applied to microarray data. J. Biopharm. Stat. **20**(2), 240–266 (2010)

R Core Team: R: A Language and Environment for Statistical Computing. R Foundation for Statistical Computing, Vienna (2018). https://www.R-project.org/. Accessed 28 Mar 2019

Richter, S.J., Richter, C.: A method for determining equivalence in industrial applications. Qual. Eng. **14**(3), 375–380 (2002)

Robinson, A.P., Froese, R.E.: Model validation using equivalence tests. Ecol. Model. **176**(3), 349–358 (2004)

Robinson, A.P., Duursma, R.A., Marshall, J.D.: A regression-based equivalence test for model validation: shifting the burden of proof. Tree Physiol. **25**(7), 903–913 (2005)

Rogers, J.L., Howard, K.I., Vessey, J.T.: Using significance tests to evaluate equivalence between two experimental groups. Psychol. Bull. **113**(3), 553 (1993)

Saavedra, M., Hellinga, B., Casello, J.: Automated quality assurance methodology for archived transit data from automatic vehicle location and passenger counting systems. Transp. Res. Record **2256**, 130–141 (2011)

Schepers, U., Wätzig, H.: Application of the equivalence test according to a concept for analytical method transfers from the International Society for Pharmaceutical Engineering (ISPE). J. Pharm. Biomed. Anal. **39**(1), 310–314 (2005)

Schuirmann, D.J.: A comparison of the two one-sided tests procedure and the power approach for assessing the equivalence of average bioavailability. J. Pharmacokinet. Pharmacodyn. **15**(6), 657–680 (1987)

Strathman, J.G.: An evaluation of automatic passenger counters: validation, sampling, and statistical inference. Technical report (1989)

Strathman, J., Kimpel, T., Callas, S.: Validation and sampling of automatic rail passenger counters for national transit database and internal reporting at TriMet. Transp. Res. Record **1927**, 217–222 (2005)

UITP: Statistics brief, urban public transport in the 21st century (2017). http://www.uitp.org/sites/default/files/cck-focus-papers-files/UITP_Statistic%20Brief_national%20PT%20stats.pdf. Accessed 28 Mar 2019

Vanhelst, J., Zunquin, G., Theunynck, D., Mikulovic, J., Bui-Xuan, G., Béghin, L.: Equivalence of accelerometer data for walking and running: treadmill versus on land. J. Sports Sci. **27**(7), 669–675 (2009)

VDV: Jahresbericht (2018). https://www.vdv.de/vdv-jahresbericht-2017-2018.pdfx. Accessed 28 Mar 2019

Verkehr & Technik: Einführung der neuen VDV 457 2.0: Automatische Fahrgastzählung der iris-GmbH erfolgreich mit max. 1% Fahrgastfehler zertifiziert (2016). https://www.VTdigital.de/VT.02.2016.062. Accessed 28 Mar 2019

VMT: Allgemeine Vorschrift für den Straßenpersonennahverkehr im VMT-Verbundgebiet (2010). https://www.vmt-thueringen.de/fileadmin/user_upload/Allgemeine_Vorschrift/Allgemeine_Vorschrift_VMT_StPNV.pdf. Accessed 28 Mar 2019

VVS: Förderrichtlinie für die Bezuschussung von automatischen Fahrgastzählsystemen in den Busverkehren der Verbundstufe II des Verkehrs- und Tarifverbunds Stuttgart (VVS) (2016). https://www.landkreis-esslingen.de/site/LRA-Esslingen-ROOT/get/params_E1561833836/13636328/87-2016%20Anlage%202%20-%20F%C3%B6rderprogramm%20AFZS.pdf. Accessed 28 Mar 2019

Waldhoer, T., Heinzl, H.: Combining difference and equivalence test results in spatial maps. Int. J. Health Geogr. **10**(1), 3 (2011)

Wellek, S.: Testing Statistical Hypotheses of Equivalence and Noninferiority. CRC Press, Boca Raton (2010)

Wellek, S., Hampel, B.: A distribution-free two-sample equivalence test allowing for tied observations. Biom. J. **41**(2), 171–186 (1999)

Westlake, W.: Bioequivalence testing—a need to rethink. Biometrics **37**(3), 589–594 (1981)

Wilson, N.H., Nuzzolo, A.: Schedule-Based Modeling of Transportation Networks: Theory and Applications, vol. 46. Springer, Berlin (2008)

Xue, R., Sun, D.J., Chen, S.: Short-term bus passenger demand prediction based on time series model and interactive multiple model approach. Discrete Dyn. Nat. Soc. (2015). https://doi.org/10.1155/2015/682390

Yi, Q., Wang, P.P., He, Y.: Reliability analysis for continuous measurements: equivalence test for agreement. Stat. Med. **27**(15), 2816–2825 (2008)

Zhao, J., Rahbee, A., Wilson, N.H.: Estimating a rail passenger trip origin-destination matrix using automatic data collection systems. Comput.-Aided Civ. Infrastruct. Eng. **22**(5), 376–387 (2007)

Zhou, J., He, Y., Yuan, Y.: Comparison of Schuirmann's two one-sided tests with nonparametric two one-sided tests. J. Pharmacokinet. **3**, 1–12 (2004)

**Michael Siebert**  is a researcher and engineer in public transport in Germany. His fundamental research in large scale railway timetabling optimization created a method which significantly improved the state of the art and whose scientific publication became a comprehensive reference in the field. Recently, by filing an expert opinion that changed the validation procedures, he became co-author of the industrial recommendation for automatic passenger counting within the VDV.

**David Ellenberger**  is a data analyst for public transport in Germany. His background comprises many years of research in the fields of mathematical statistics, life sciences, and public transport with a particular focus on high-dimensional data, evidence synthesis, survey methodology, machine learning, and passenger modelling. Recent contributions in the field of automatic passenger counting have been recognized by the VDV, and have led to co-authoring of the industrial recommendation.